# Seismic isolation of Advanced LIGO:
# Review of strategy, instrumentation and performance

F. Matichard[1,2,*], B. Lantz[3], R. Mittleman[1], K. Mason[1], J. Kissel[5], J. McIver[7], B. Abbott[2], R. Abbott[2], S. Abbott[2], E. Allwine[5], S. Barnum[1], J. Birch[4], S. Biscans[1,2], C. Celerier[3], D. Clark[3], D. Coyne[2], D. DeBra[3], R. DeRosa[6], M. Evans[1], S. Foley[1], P. Fritschel[1], J.A. Giaime[4,6], C. Gray[5], G. Grabeel[5], J. Hanson[4], C. Hardham[3], M. Hillard[1], W. Hua[3], C. Kucharczyk[3], M. Landry[5], A. Le Roux[4], V. Lhuillier[5], D. Macleod[6], M. Macinnis[1], R.Mitchell[4], B. O'Reilly[4], D. Ottaway[1], H. Paris[3,5], A.Pele[4], M. Puma[4], H. Radkins[5], C. Ramet[4], M. Robinson[5], L.Ruet[1], P. Sarin[1], D. Shoemaker[1], A. Stein[1], J. Thomas[4], M. Vargas[4], K Venkateswara[8], J. Warner[5], S.Wen[3]

[1]MIT, Cambridge, MA, USA. [2]Caltech, Pasadena, CA, USA, [3]Stanford University, Stanford, CA, USA, [4]LIGO Livingston Observatory, Livingston, LA, USA, [5]LIGO Hanford Observatory, Hanford, WA, USA, [6]Louisiana State University, Baton Rouge, LA, USA. [7]University of Massachusetts, Amherst, MA, USA. [8]University of Washington, Seattle, WA, USA

**Abstract**

The new generation of gravitational waves detectors require unprecedented levels of isolation from seismic noise. This article reviews the seismic isolation strategy and instrumentation developed for the Advanced LIGO observatories. It summarizes over a decade of research on active inertial isolation and shows the performance recently achieved at the Advanced LIGO observatories. The paper emphasizes the scientific and technical challenges of this endeavor and how they have been addressed. An overview of the isolation strategy is given. It combines multiple layers of passive and active inertial isolation to provide suitable rejection of seismic noise at all frequencies. A detailed presentation of the three active platforms that have been developed is given. They are the hydraulic pre-isolator, the single stage internal isolator and the two-stage internal isolator. The architecture, instrumentation, control scheme and isolation results are presented for each of the three systems. Results show that the seismic isolation sub-system meets Advanced LIGO stringent requirements and robustly supports the operation of the two detectors.

## 1 Introduction

To perform the first direct detection of gravitational waves, new generations of detectors require unprecedented levels of seismic isolation. This paper summarizes the results of over a decade of research on active inertial seismic isolation for the Advanced LIGO detectors. The goals of this article are to review the strategy chosen to meet these stringent requirements, to describe the instrumentation developed to support this strategy, and to summarize the performance achieved at the two LIGO observatories.

Active inertial isolation strategies for physics experiments has been investigated in the 1980s and 1990s for high-accuracy measurements of the earth's gravitational field [1-3]. This work led the way to a new generation of platform actively controlled using inertial instruments. In 1991, Nelson presented results on an isolation stage servo-controlled in the vertical direction with inertial sensors [4]. He pointed out the factors limiting the performance of such systems, namely the internal mechanical resonances reducing the control bandwidth, the sensor noise limiting the isolation, and the cross couplings inducing drive and sensing errors. Soon after, work on multi-stage active platforms emerged to provide further isolation. In 1994, Stebbins et al. published experimental results for a multi-stage platform [5]. In 1998, Richman et al. presented results for the second of a three-stage system [6].

The first generation of LIGO gravitational wave observatories (Initial LIGO) was being built in the 1990s [7]. Saulson outlined vibration isolation concepts for those detectors and discussed the parameters limiting the performance [8]. While active isolation solutions were proposed for early LIGO experiments [9], passive stacks were preferred for Initial LIGO in order to support and isolate the suspended mirrors [10]. During the following years, a proposal was written to upgrade the LIGO observatory with a much more sensitive detector called Advanced LIGO [11]. This new generation required very demanding levels of seismic isolation [12]. The approach chosen for Advanced LIGO was to develop the active platforms and suspensions independently to facilitate the design, construction and commissioning. Large, heavy and stiff active platforms would provide the low frequency isolation [13]. Soft suspensions mounted on the platforms would provide several stages of passive isolation to the interferometer mirrors [14].

In 2001, Giamie et. al. presented a conceptual design of active platforms for Advanced LIGO [15]. The proposal was based on the Joint Institute for Laboratory Astrophysics (JILA) experience [5, 6], the work done at Stanford on quiet hydraulic actuation [16], active isolation techniques [17], and experimental results obtained with a Rapid Prototype [15]. This original concept was designed to meet Advanced LIGO's stringent requirements in terms of isolation, positioning, alignment, payload capacity and vacuum compatibility. The concept was refined in the following years [18-20]. This led to the design of the three platforms used for the Advanced LIGO project. At each observatory, eleven hydraulic platforms called HEPIs are used as pre-isolation stages externally to the vacuum chamber. Five single-stage Internal Seismic Isolators are mounted on the HEPI platforms and located in the vacuum chambers called the Horizontal Access Modules (HAM). They support and isolate the interferometer's auxiliary optics from ground motion, and are called the HAM-ISI platforms. Five in-vacuum two-stage seismic isolators are mounted on the HEPI platforms and located in the vacuum chambers called the Basic Symmetric Chambers (BSC). They isolate the interferometer's core optics from ground motion and are called the BSC-ISI platforms.

*fabrice.matichard@ligo.org

This paper reviews these three Advanced LIGO active seismic isolation platforms. It summarizes the history, development, lessons learned, and performance achieved, from the early stage of development to date. The second section gives an overview of the Advanced LIGO system. The third section gives an overview of the isolation strategy, and the controls infra-structure developed to support this approach. The fourth section reviews the hydraulic stage of pre-isolation, the fifth section reviews the single stage platform, and the sixth section reviews the two-stage platform.

## 2 System overview

Advanced LIGO observatory is designed to perform the first direct detection of gravitational waves [21]. It consists of two 4 km interferometric detectors installed at the Hanford (WA) and Livingston (LA) sites. A representation of the vacuum envelope and equipment hosting the instruments at the Livingstion site is shown in Fig. 1. Each detector uses 11 vacuum tanks. Five of them are large BSC chambers (approximately 4.5 meters high, and 2.5 meters in diameter) housing the interferometer's core optics. Six of them are the smaller HAM chambers (approximately 2.5 meters high, and 2.5 meters width) housing the interferometer's auxiliary optics.

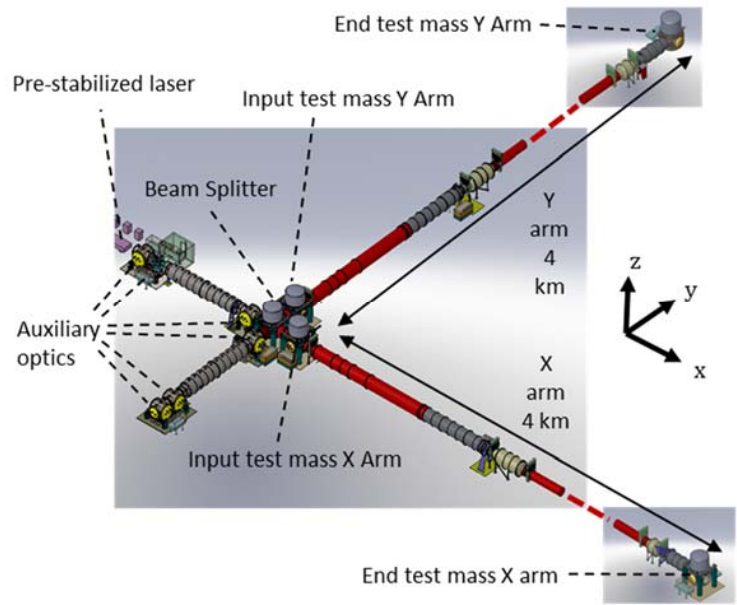

**Fig. 1: Representation of the LIGO vacuum system.**

In the HAM chambers, three systems are cascaded to provide up to five stages of seismic isolation to each auxiliary optic. A conceptual and a Computer Aided Design (CAD) model are shown in Fig. 2. The first system is the Hydraulic External Pre-Isolator (HEPI). This active platform provides one stage of active isolation. The second system is the Internal Seismic Isolation platform for HAM chambers (HAM-ISI). This system provides one stage of isolation combining active and passive isolation. The auxiliary optics are mounted on the HAM-ISI table. Most auxiliary optics components are suspended by passive suspensions. Up to the three stages of suspensions, inspired by GEO 600 suspensions [22], are used for the auxiliary optics.

In the BSC chambers, three systems are cascaded to provide up to seven stages of seismic isolation to the core optics. A conceptual and a CAD model are shown in Fig. 3. As previously, the HEPI active platform provides the first stage of isolation. The second system is the Internal Seismic Isolation platform for BSC chambers (BSC-ISI). It provides two stages of isolation, each combining both active and passive isolation. The core optics are mounted on the down-facing BSC-ISI optical table. The test masses are suspended at the last stage of quadruple pendulum suspensions [23].

In this approach, the HEPI platform using hydraulic actuators provides a long range positioning and alignment capability (on the order of a millimeter). The Internal Seismic Isolation platforms (HAM-ISI and BSC-ISI) include optical tables on which are mounted the interferometer components. The ISI systems use low noise inertial sensors to provide low frequency active isolation (as low as 0.1 Hz). The suspensions mounted on the ISI platforms cascade several stages to provide the passive isolation necessary to attenuate the seismic motion to adequate levels in the observational bandwidth (above 10 Hz). The HEPI platform, the HAM-ISI platform, and the two stages of the BSC-ISI platform use different architecture and instrumentation, but they share similar active isolation principle. The next section summarizes the isolation and control principle of these active platforms.

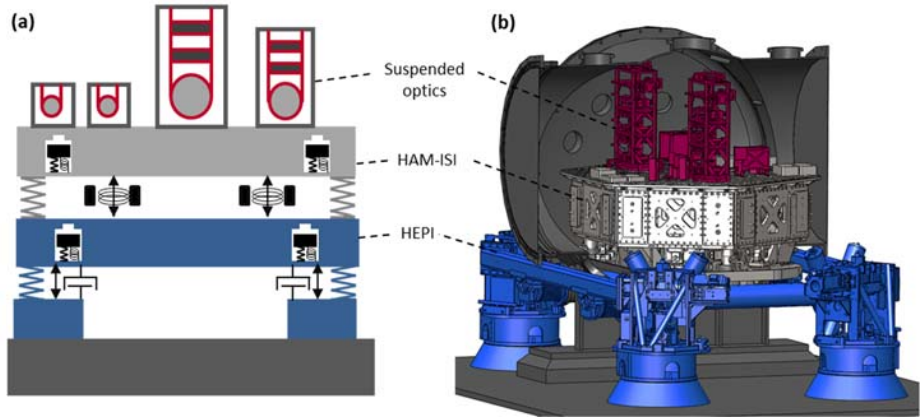

**Fig. 2: (a) schematic and (b) CAD model of the isolation systems supporting the auxiliary optics in the HAM chambers.**

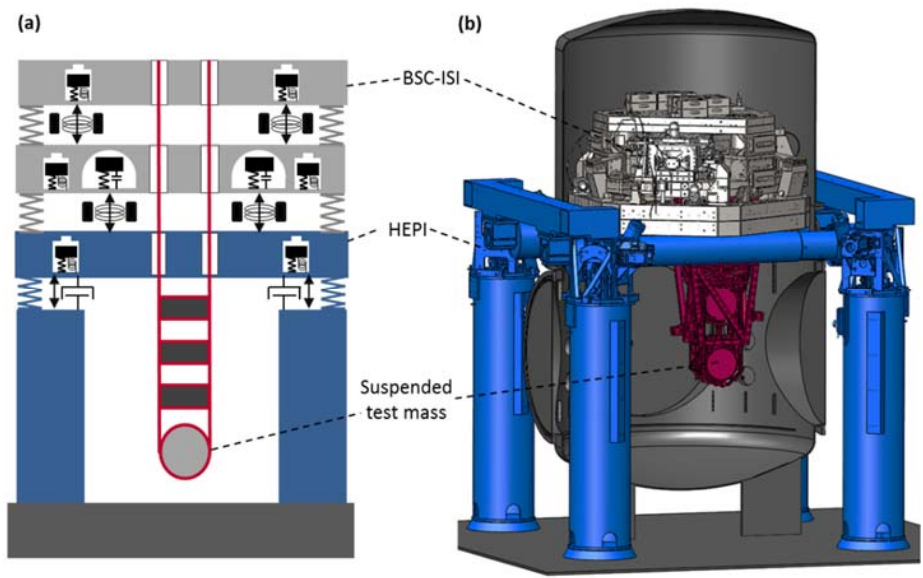

**Fig. 3: (a) schematic and (b) CAD model of the isolation systems supporting the core optics in the BSC chambers.**

## 3 Inertial isolation scheme and control infra-structure

### 3.1 Isolation and control overview

The passive-active concept used in Advanced LIGO isolation platforms can be summarized by the schematic in Fig. 4. The motion disturbance transmitted by the support structure (or the previous isolation stage) is shown in grey (0). The isolation platform (1) is supported by suspension springs (2). Above the resonance frequency, the platform is inertially decoupled from the input stage and provides passive isolation. Relative sensors (3) are used to servo-position the platform with respect to the support structure at very-low frequencies. Inertial sensors (4) are used to provide active inertial isolation through feedback control. The signals from all the sensors are combined in a sensor fusion to drive the control forces (5). Additional performance is obtained using feedforward inertial sensors (6). The platforms are designed to minimize the cross couplings between the degrees of freedom (DOF). Each of the six DOF can be controlled using independent single input single output control loops.

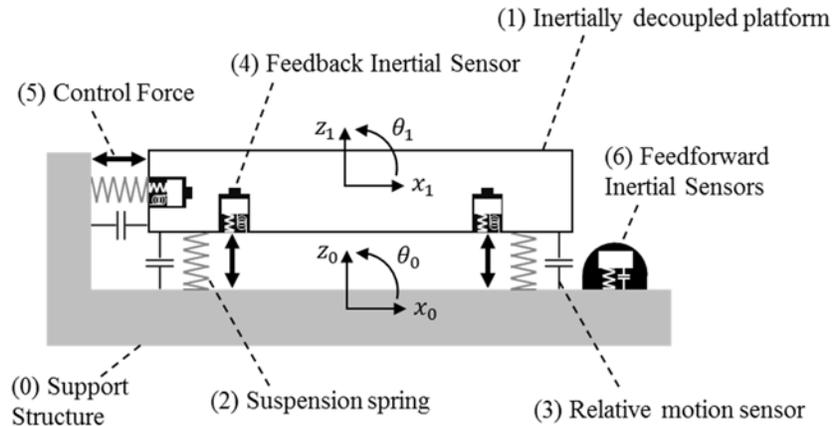

**Fig. 4: Schematic representation of a passive-active inertial isolation system**

Each platform (or stage of a platform) is equipped with a set of relative position sensors, a set of inertial sensors (geophones and/or seismometers), and a set of actuators. A minimum of six instruments for each set of sensors (or each set of actuators) is necessary to sense (or drive) of all the rigid body motions. More sensors can be used and combined in a sensor blend/fusion to reduce the sensor noise, or to sense deformation modes. The local measurements (or actuations) performed by each individual sensor (or actuator) are symbolically represented with vectors in Fig. 5 (a). In this example, there are three horizontal instruments (sensors or actuators) mounted tangentially, and three vertical instruments.

The local relative sensor signals are combined in real time to measure the platform's relative displacement along the Cartesian basis (axis of the interferometer), as illustrated on the right of Fig. 5 (b). A matrix based on the geometry (instruments location and orientation) is used to perform the change of basis. Details of the matrix calculations can be found in [24]. A similar construction is used to sense the absolute motion of the platform in the Cartesian basis using the set of inertial sensors. The local actuators are combined to drive the platform with forces and torques in the Cartesian basis. All the controls are done in the Cartesian basis. The rigid body translations and rotations (called X, Y, Z, RX, RY, RZ) are controlled independently of each other. The next section describes the topology used to control each of the independent DOF.

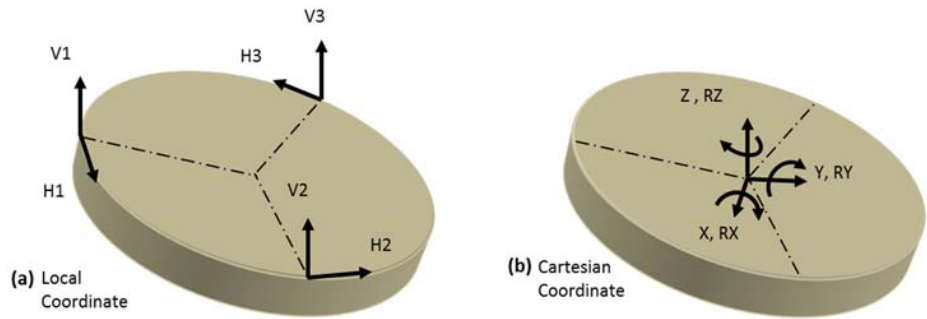

**Fig. 5: Sensing and driving change of basis**

### 3.2  Control scheme

The block diagram in Fig. 6 shows the generic topology used to control one degree of freedom (X, in this example). The seismic path (transmissibility) is denoted $P_s$, and the actuator path (force or torque transfer function) is denoted $P_F$. The control force F is used to reduce the platform motion denoted $X_1$ induced by the input motion disturbance denoted $X_0$. The control scheme includes three main components: a feedback block, a sensor correction block and a feedforward block.

The platform motion $X_1$ is the sum of the input disturbance and the control force contributions as shown in equation (1). The feedback block is the central part of the control scheme. The measurement of the platform's absolute motion made with the inertial sensors is called $U_a$. The sensor noise of the inertial instruments is $N_a$. Only inertial sensors (accelerometers, geophones, seismometers) can be used to provide seismic isolation, but such instruments are inherently AC coupled and they cannot be used at very low frequencies (noise limited). For that reason, relative sensors are used at low frequencies to measure the differential motion between the output and the input stages ($X_1 - X_0$). This sensor cannot be used to provide seismic isolation, but it allows low frequency positioning with respect to the input stage. The relative sensor signal is denoted $U_r$. The noise of this sensor is denoted $N_r$.

The relative motion signal $U_r$ is low passed with the filter $L$, and the absolute motion signal $U_a$ is high passed with the filter $H$. The filters $H$ and $L$ are tuned to obtain a suitable compromise between active seismic isolation at "high" frequencies (typically above 100 mHz), and motion amplification induced by sensor noise at low frequencies (below 100 mHz) [25]. The design of $H$ and $L$ is based on complementary filters ($L + H = 1$ for all frequencies). The frequency at which the inertial sensor and relative sensor have equal participation is called the blend frequency. The signal resulting from this blend (sensor fusion) is called the super sensor. It feeds the feedback controller called $C_{FB}$.

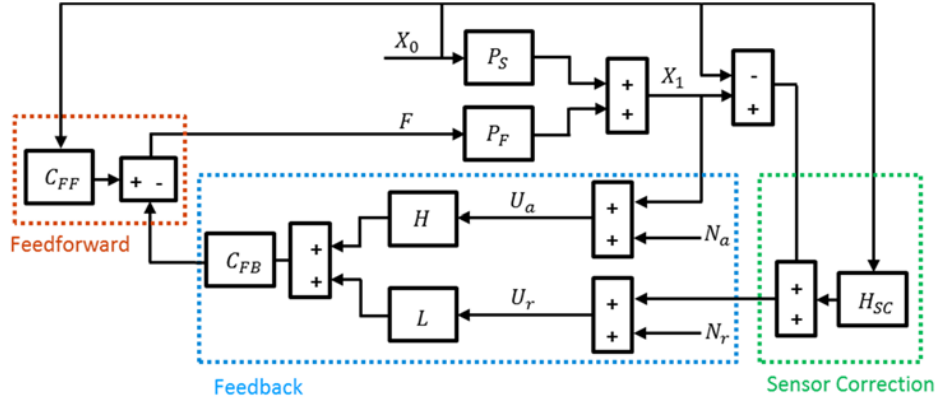

**Fig. 6: Simplified block diagram of the active control scheme**

Equation (2) gives the feedback control force as a function of the blended signal. Using complementary filters as written in Eq.(3), the closed loop transmissibility simplifies to the expression given in Eq. (4). The denominator shows the loop gain attenuation. The numerator shows that the transmissibility is increased by the term ($L\,C_{FB}\,P_F$) induced by the relative sensor control. The limit given in Eq. (5) shows that in the control bandwidth (where the loop gain is high), the transmissibility tends to the low pass filter applied to the relative sensor.

The time average of the noise contributions is expressed in Eq. (6). This equation gives the output stage power spectral density assuming the input motion, inertial sensor noise and relative sensor noise are uncorrelated. Eq. (7) gives the limit of the amplitude spectral density for high loop gain. This expression shows the influence of the low pass filter $L$ on the transmissibility and on the relative sensor noise injection. It shows the influence of the high pass filter $H$ on the inertial sensor noise injection. The low pass filter (and its complementary high pass), are tuned to obtain a suitable compromise between isolation above the blend frequency and noise injection below the blend frequency.

$$X_1 = P_s\,X_0 + P_F\,F \tag{1}$$

$$F = -C_{FB}\,(H\,X_1 + L\,(X_1 - X_0)) \tag{2}$$

$$H + L = 1 \tag{3}$$

$$\frac{X_1}{X_0} = \frac{P_s + L\,C_{FB}\,P_F}{1 + C_{FB}\,P_F} \tag{4}$$

$$\lim_{(C_{FB}\,P_F)\to\infty}\left(\frac{X_1}{X_0}\right) = L \tag{5}$$

$$\langle X_1^2 \rangle = \left|\frac{P_s + L\,C_{FB}\,P_F}{1 + C_{FB}\,P_F}\right|^2 \langle X_0^2 \rangle + \left|\frac{H\,C_{FB}\,P_F}{1 + C_{FB}\,P_F}\right|^2 \langle N_a^2 \rangle + \left|\frac{L\,C_{FB}\,P_F}{1 + C_{FB}\,P_F}\right|^2 \langle N_r^2 \rangle \tag{6}$$

$$\lim_{(C_{FB}\,P_F)\to\infty} \sqrt{\langle X_1^2 \rangle} = \sqrt{\langle (L\,X_0)^2 \rangle + \langle (H\,N_a)^2 \rangle + \langle (L\,N_r)^2 \rangle} \tag{7}$$

An example of complementary filters is given in Fig. 7 (a). In this example, the blend (cross-over between $N$ ) frequency is 200 mHz. The super sensor signal is dominated by the relative sensor signal below 200 mHz and by the inertial sensor signal above that frequency. The high-pass filter is designed at low frequency to sufficiently filter the inertial sensor noise $N_a$. In the control bandwidth (up to 25 Hz), the low pass filter is designed to provide adequate isolation. Notches have been added to reduce the transmissibility at frequencies of particular interest (1 Hz and 1.5 Hz, which correspond to the resonances of the suspensions mounted on the platform). At higher frequency, the low pass filter is designed so that the relative sensor noise does not compromise the isolation. In the control bandwidth, the transmissibility follows the low pass. Additional isolation can be obtained using higher order filtering in the low pass filter. This would however result in higher motion amplification near the blend frequency.

For practical implementation, both signals are calibrated in the same units in order to be combined with the complementary filters. The inertial sensor signal is integrated ("stretched") before entering the high pass filter. Calibration and complementary filters are usually combined for practical implementation in the digital real time system.

The second block of the control scheme is called sensor correction. It is shown in the block diagram in Fig. 6. It uses an inertial instrument sensing the ground motion. This signal is filtered and added to the relative sensor signal as shown in Eq. (8). The inertial sensor provides the information needed to reduce the transmissibility that is limited by the relative sensor part of the super sensor, as written in Eq. (9).

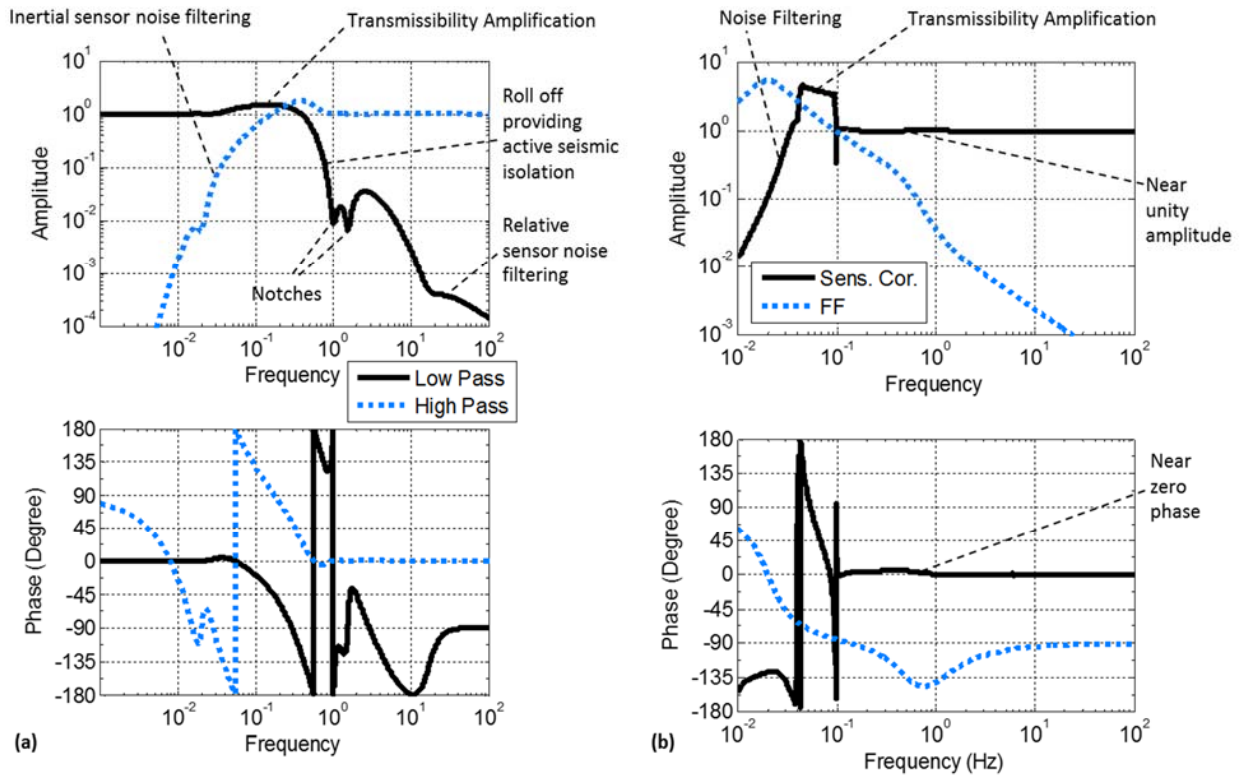

**Fig. 7: (a) example of complementary filters and (b) FIR filter and feedforward filters [25]**

The isolation that can be obtained through sensor correction depends on the coherence between the witness sensor (ground instrument) and the platform motion ($X_1$). One benefit of the sensor correction technique is that a single low noise instrument can be used for the sensor correction of multiple platforms. Another benefit is that the sensor noise injected is common to all the platforms fed by the same ground instrument. Therefore, the low frequency motion amplification possibly caused by sensor noise injection is common to all the platforms. This is a key feature for interferometric cavities sensing the differential motion between optics mounted on different platforms.

The limit in Eq. (10) shows the transmissibility for high loop gain, assuming perfect coherence between the ground instrument and the platform motion. A sensor correction filter $H_{sc}$ equal to unity at all frequencies would cancel the transmissibility. Another way to phrase this is that the inertial instrument sensing $X_0$, added to the relative sensor ($X_1 - X_0$), provides an ideal measurement of the absolute motion $X_1$. The ground inertial instrument is nevertheless noise limited at low frequency. Therefore it must be adequately filtered by $H_{sc}$. Fig. 7 (b) shows an example of a sensor correction filter. It is a Finite Impulse Response (FIR) filter designed by Hua [25]. It provides near perfect amplitude and phase matching above 100 mHz where most isolation is needed. At those frequencies, this filter minimizes the residual motion given in Eq. (10). The filter provides high pass filtering at very low frequency to not inject sensor noise, at the cost of motion amplification between 50 mHz and 100 mHz. This filter provides a compromise accounting for ground motion spectra (peaking at the micro-seismic motion above 100 mHZ), and the inertial sensor noise dominating the signal at low frequencies.

$$U_r = (X_1 - X_0) + H_{SC}\, X_0 \tag{8}$$

$$\frac{X_1}{X_0} = \frac{P_s + L\, C_{FB}\, P_F\, (1 - H_{SC})}{1 + C_{FB}\, P_F} \tag{9}$$

$$\lim_{(C_{FB}\, P_F) \to \infty} \left(\frac{X_1}{X_0}\right) = L(1 - H_{SC}) \tag{10}$$

The sensor correction operates only where the loop gain is sufficiently high as shown in Eqs. (9) and (10). If residual coherence exists between a witness instrument and the platform motion, a standard feedforward control can be implemented to reduce further the transmissibility as shown in the feedforward block diagram in Fig. 6. Equation (11) gives the definition of the ideal feedforward controller accounting for the sensor correction already implemented. The filter is typically band-passed so as not to deteriorate the transmissibility at frequencies where the coherence is low. An example of such a feedforward filter is shown in Fig. 7 (b).

$$C_{FF} = -\frac{P_s}{P_f} - L\, C_{FB}\, (1 - H_{SC}) \tag{11}$$

### 3.3 Controls infrastructure

Fig. 8 (a) gives a schematic overview of the electronics diagram of a seismic isolation platform. Actual electronics are shown in Fig. 8 (b). Custom low noise electronics boards and chassis have been designed to condition the instruments signal and to distribute them to the Analog to Digital Converter (ADC), to receive the drive signals from the Digital to Analog Converter (DAC), and then to condition and distribute them to the actuators.

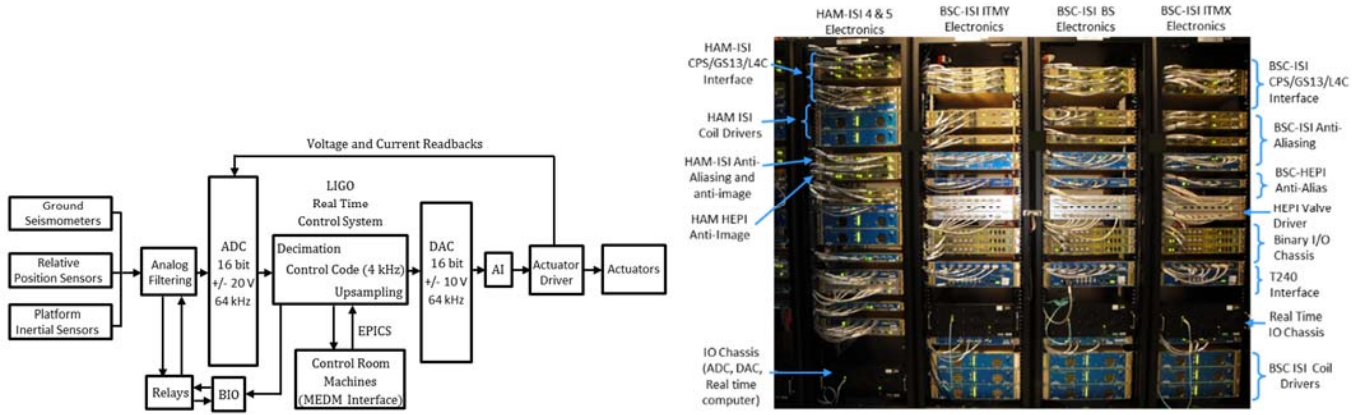

**Fig. 8: Diagram (a) and actual picture (b) of the isolation systems electronics.**

The detailed design is described in [26]. It includes low noise pre-amplification boards designed for the passive geophones [27]. Their design is based on noise budget estimates of the platform motions as described in [28]. Amplification and whitening filter stages are used to maintain the signals above the ADC noise. These stages of amplification can be switched remotely as a function of operating conditions via binary IO cards and relays. The electronics provide read back signals to monitor the state of the switchable filters. The output electronics include anti-image and actuator drivers (valve drivers of hydraulic actuators and coil drivers of electromagnetic actuators). These filtering stages are designed to reduce the DAC noise transmitted to the drivers. Coil drivers voltage and current read backs are sent to the ADC for monitoring.

The signals are digitized at 64 kHz using 16 bits +/- 20V ADC cards. A third order Chebyshev at 10 kHz with a notch at $2^{16}$ Hz is used for the anti-aliasing filter. The signals are decimated to the real time controller sampling frequency (at 2kHz for HEPI platforms, 4 kHz for HAM-ISI and BSC-ISI). The controller output is up-sampled to the DAC cards frequency at 64 kHz. The digital controller is based on the LIGO CDS real time code [29]. An Epics database [30] is used for communication with the control room machines. The operator interface is built in MEDM code [31]. A software system was developed to coordinate and automate the hierarchical and sequential turn on and turn off process of Advanced LIGO subs-systems (just for the seismic isolation system, each detector uses 21 active platforms, which include up to 12 DOF per platform, several hundreds of parallel feedback and feedforward loops, and thousands of digital channels) [32].

# 4 The hydraulic external pre-isolator (HEPI)

## 4.1 HEPI Overview

All Advanced LIGO chambers are equipped with Hydraulic External Pre-Isolation systems. The HEPI platform combines quiet hydraulic actuators, inductive position sensors, geophones and ground seismometers to provide very low frequency active isolation. While this concept was initially proposed for Advanced LIGO, its development was accelerated to provide isolation at the LIGO Livingston Observatory during Initial LIGO. The Livingston observatory location, even though carefully studied and chosen [33], was suffering from anthropogenic motion causing substantial differential motion over the 4 km baseline of the interferometer [34]. The external driver based on actuated piezo-electric stacks used during initial LIGO did not have enough control authority to maintain the interferometer locked over long stretches of time. It was decided to accelerate the design and installation of a new external pre-isolator. Two actuation techniques were studied and prototyped. A version called MEPI using electromagnetic actuators and a second one using hydraulic actuators called HEPI. Hardham discussed the advantages and drawbacks of the two techniques, and compared transfer functions obtained with the two prototypes [35]. The hydraulic version was selected for its long range of motion and its inherent viscous effect that damps the structural modes.

The hydraulic actuator used in the HEPI system is based on a quiet hydraulic drive technique. This field was investigated by DeBra at Stanford University [16]. In such actuators, the low noise performance is achieved by operating entirely in a laminar flow regime, and by the absence of frictional interfaces. The initial design, prototyping, testing and final design of the actuators used in HEPI was presented by Hardham [36]. After successful tests carried out at the LIGO-MIT test facilities [37], the HEPI system was installed at the LIGO Livingston observatory. Used to reject tidal, microseism and anthropogenic motion, it significantly increased the interferometer's duty cycle of operation. Hua's work on high-pass digital filters, implemented as a polyphase filter, contributed to the degree of isolation achieved with the system [25, 38]. These high-pass digital filters used for sensor correction provided near perfect amplitude and phase matching in the frequency band of interest (at the ground's second microseism peak), at the cost of motion amplification at frequencies of lesser interest (below the microseism). Wen presented detailed isolation results from measurements made on the HEPI system installed at the Livingston Observatory for initial LIGO [39, 40]. He demonstrated significant isolation performance in the control bandwidth. For the horizontal directions, he measured the motion amplification at low frequencies, and correlated it with the tilt-horizontal coupling inherent to the use of inertial sensors. It was concluded that a tilt sensor would be needed to obtain further isolation through sensor correction, which led Lantz et al. to define requirements for a ground tilt sensor [41]. The HEPI actuation capability has also been used to suppress residual coherent motion between ground instruments and the interferometer signal in a feed forward scheme [42].

During the past years, the Advanced LIGO HEPI platforms have been installed and tested at the two LIGO observatories. This configuration integrates a stiffer frame for the HAM chambers, and a control scheme refined to work in conjunction with the Advanced LIGO payload it supports. The next section summarizes the HEPI instrumentation. It explains how the lessons learned from the initial experiments have been incorporated to enhance the structural behavior and isolation performance.

## 4.2 HEPI architecture and instrumentation

The HEPI system has been designed to provide very low frequency active isolation and precision positioning. There are two HEPI configurations: one for the auxiliary optics chambers and one for the core optics chambers. A conceptual representation, a CAD model and a picture of the of the HEPI system for the HAM chambers are shown in Fig. 9.

The sensor and actuator instrumentation is common to both configurations. Only the payload, the suspended frame and the structure geometry vary between the two configurations. The suspended structure is a quadrilateral rigid frame made of two support tubes and two cross beams as shown in Fig. 9 (a) and (b). It is suspended in each of the four corners by a supporting structure called the HEPI housing. Each of the four housing structures includes two maraging steel helical springs mounted in a v-shape, two inductive position sensors for low frequency positioning, two geophones for inertial isolation and two hydraulic actuators as shown in Fig. 10 (a). Bellows are used at the interface between the air and the vacuum. In-vacuum equipment (a HAM-ISI or BSC-ISI) is bolted to the support tubes.

The springs are mounted with a 30 degree angle with respect to the vertical axis. Each spring is made of double-start helicals that are counter-wound to reduce the vertical to rotational motion coupling. Maraging steel 300 is used for its high strength (2015 MPa yield point). For the heaviest payload configuration (core optics chambers), the mass $m$ of the load carried by HEPI is about 6500 kg.

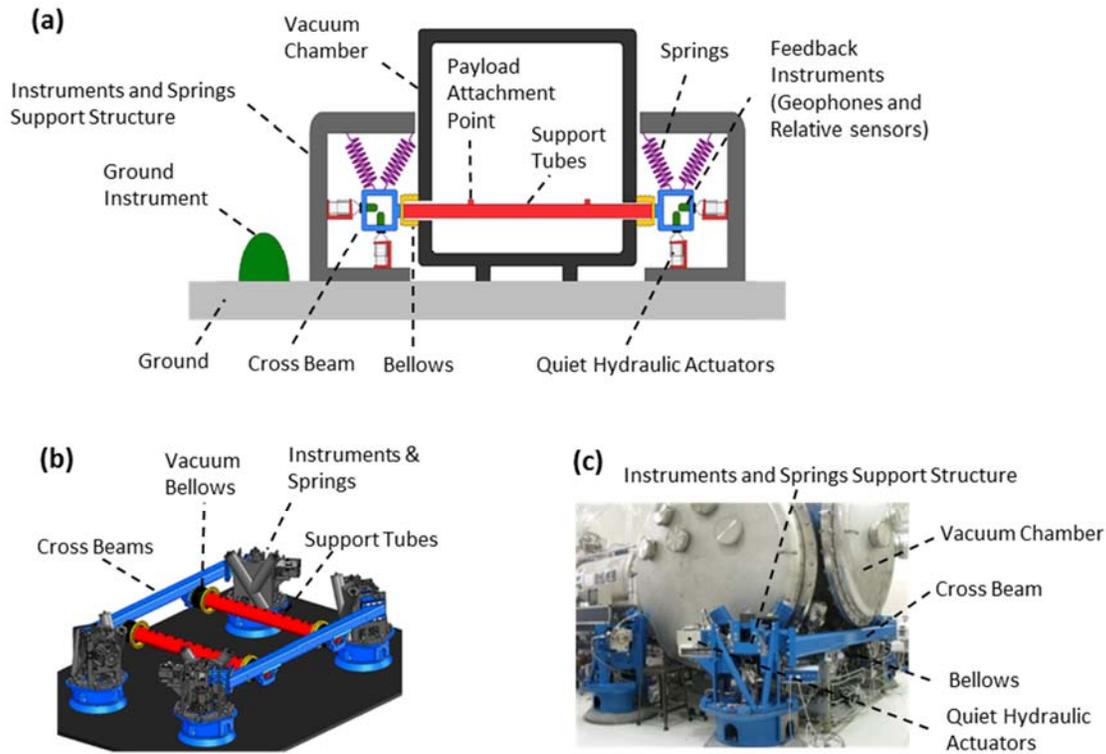

**Fig. 9: (a) Conceptual representation, (b) CAD and (c) picture of the of the HEPI system for the HAM chambers.**

Quiet hydraulic actuators, shown in Fig. 10 (b), are used to drive the HEPI platforms [36]. While conventional hydraulic actuators operate in the turbulent flow regime resulting in pressure fluctuations and actuation noise, the HEPI actuator operates in the laminar flow regime to reduce the actuator noise. The differential pressure between the two chambers of the actuator drives the tripod connected to the structure. It is controlled with a servo-valve made of a bridge of hydraulic resistances (hydraulic Wheatstone bridge configuration). Flexible bellows are used instead of a piston to suppress friction at the interface between the moving parts. The bellows have a convoluted shape designed to maximize the ratio of breathing stiffness to axial compliance. A bypass network is included in the design to damp the breathing resonance in the bellows. The actuator also includes a bleed network to remove entrapped air. A tripod is used at the interface with the structure to reduce the transverse stiffness. Besides low noise actuation, the actuator viscous properties are also useful in damping the rigid body modes which makes the system robust. The deformation modes of the frames are also damped which simplifies the design of the feedback control filters. The next section summarizes the control and experimental results of the HEPI system.

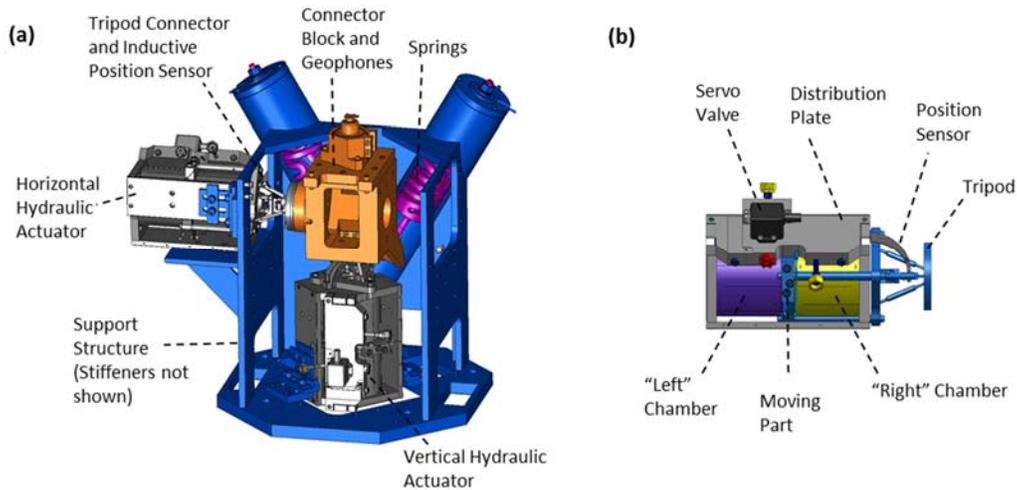

**Fig. 10: (a) HEPI spring and instruments housing, and (b) HEPI quiet hydraulic actuators.**

### 4.3 HEPI Control and experimental results

The individual HEPI actuator forces are combined to drive the platform in the Cartesian basis, and the sensors are combined to sense the platform motion in the same basis, as described in section 3. Fig. 11 (a) shows examples of HEPI transfer functions for a HAM configuration. The solid black curve is the transfer function in the X direction measured by the inductive position sensors (IPS). The first pole of the transfer function is near 20 mHz. Above that frequency, the response falls as $1/f$. The hydraulic dashpot effects damp all the rigid body and structural resonances, which greatly simplifies the design of the control filters. The support structure deformation induces the *zeros* at 20 Hz and 50 Hz.

The dotted blue line shows the transfer function in the X direction measured by the geophones (L4C). The response is calibrated in displacement units. This curve shows a corner frequency at 200 mHz due to the geophone sensitivity to tilt horizontal coupling. Below this frequency, the signal is dominated by tilt [41]. Above this frequency, the signal is dominated by translation. The tilt sensed by the geophone is a combination of global rotation (misalignment between the actuator plane and the static center of rotation) and local deformation of the frame. For Advanced LIGO, the frame has been redesigned to be stiffer [43]. Characterization results for the new design were presented in [44]. The solid and dashed curves in Fig. 11 (a) show the transfer functions in the vertical direction measured by the IPS and L4C sensors respectively. The Y transfer functions are similar to those for X. The RX, RY and RZ transfer functions are similar to those of Z. There are 8 actuators and sensors, for 6 rigid body modes, therefore two deformation modes can also be driven, sensed and controlled.

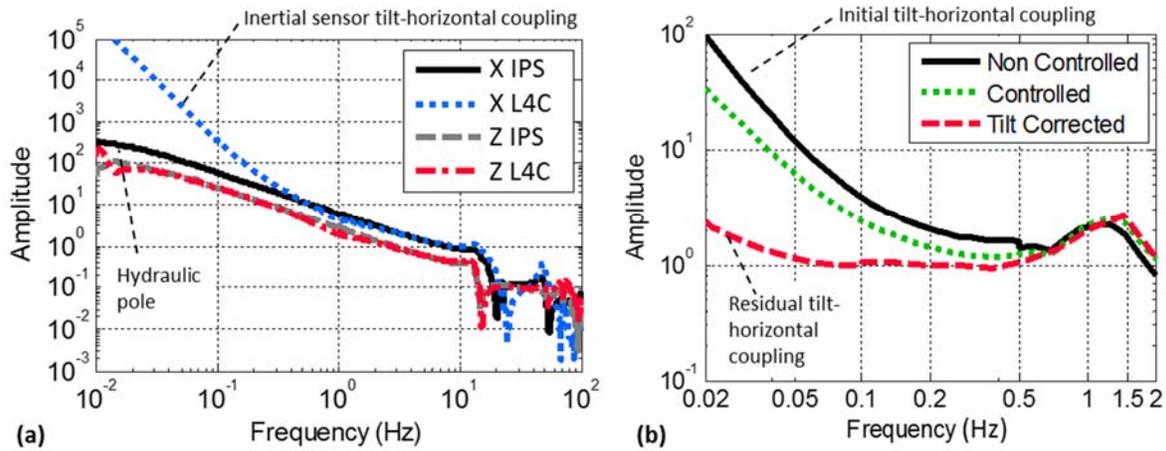

**Fig. 11: (a) HEPI transfer functions and (b) tilt decoupling.**

The tilt-horizontal coupling induced by the horizontal drive can be quantified and reduced as shown in Fig. 11 (b). In the three transfer functions shown, the HEPI actuators are used to drive the system in the X direction. The X motion is measured with the horizontal inertial sensors of the HAM-ISI which are good witnesses of the global tilt motion (better than the HEPI L4C geophones which rotates as the HEPI frame bends). The solid curve shows the open loop response (HEPI not controlled). Tilt motion dominates the response below 200 mHz. The dotted line shows the response with the HEPI control loops closed. The vertical loops constrains the global rotations, and the tilt frequency reduces to about 100 mHz. To suppress the residual tilt, a torque proportional to the translation is applied through a feedforward scheme. This correction further reduces the tilt-horizontal coupling as shown by the dashed curve. The transfer function is now flat down to 50 mHz, showing that the signal is dominated by translation down to that frequency. Limitations to tilt decoupling and tilt subtraction are discussed in Ref. [45].

The HEPI system is controlled in the Cartesian basis as described in section 3. The control topology for one degree of freedom is summarized in Fig. 12 (a), using the *X* direction as an example. The ground motion $X_0$ drives the platform motion $X_1$ through the seismic path $P_s$. A force in the *X* direction is used to control the platform through the force path $P_f$. The geophones signals are combined in the *L4C* block to sense the absolute motion in the *X* direction. The inductive position sensors signals are combined in the *IPS* block to sense the relative motion in the *X* direction. The inertial sensor signal is calibrated in displacement units and high passed with the filter *H* designed to filter the sensor noise at low frequency. The relative sensor signal is low-passed with the filter *L* designed to provide adequate seismic isolation. The filters *H* and *L* are based on complementary filters with crossover frequency typically around 0.7 Hz.

An example of a controller is shown in Fig. 12 (b) for the X direction. The transfer function ($P_f$) is shown by the solid line, the compensator is shown with the dotted line, and the open loop transfer function is shown with the dashed line. It provides a loop gain of $10^3$ at 100 mHz and over $10^2$ at 1 Hz. This control loop is conditionally stable in order to provide high loop gain in the control bandwidth. The unity gain frequency is set near 10 Hz with a phase margin of 50 degrees. Additionally, a low-noise long-period instrument is used for sensor correction. To accomplish this, the filter $H_{sc}$ calibrates the seismometer's signal into displacement units and high passes it.

An example of isolation results is presented in Fig. 12 (c), for a BSC unit. The ground motion sensed with a long-period seismometer is shown by the solid black curve. The motion of HEPI with the control off is shown by the dotted line (as measured by the in-loop L4C geophones). At most frequencies, HEPI is viscously coupled the ground. Structural resonances of the support structure induce the features at 9 Hz [46]. The sensor correction and inertial feedback combine to provide active isolation from about 0.1 Hz to 8 Hz. Up to 40 dB of isolation is achieved in the control bandwidth, where it is the most needed for the commissioning and operation of the Advanced LIGO interferometer. Similar performance is obtained in the vertical direction. Less isolation is obtained for the rotational DOF as no sensor correction is implemented in rotation. The pre-isolator provides robust and quiet support for the HAM-ISI and BSC-ISI systems which are presented in the next sections.

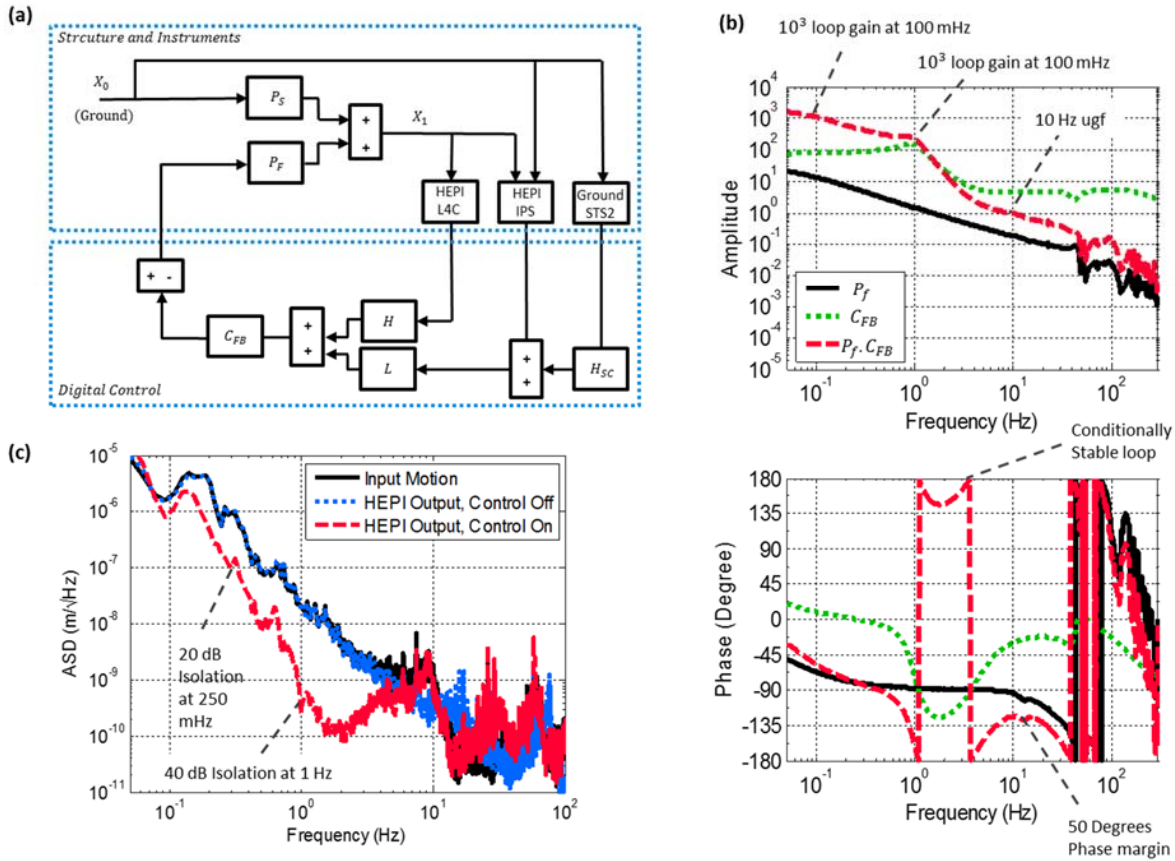

**Fig. 12: (a) HEPI control diagram, (b) HEPI controller, and (c) isolation results.**

## 5 The single stage internal isolator (HAM-ISI)

**5.1 HAM-ISI overview**

In 1999, Fritschel et al presented the initial set of requirements for the auxiliary optics chambers [47]. They set the specification on dimensions (1.9 m x 1.7 m), loading capacity (to supporting an 800 kg of payload), structural stiffness (no internal resonances below 100 Hz), and isolation performance. At the time, it was envisioned that two stages of isolation would have been necessary to meet the seismic motion specifications of auxiliary optics. The requirements were relaxed in the following years [48], thanks to the progress made on other sub-systems [49]. Soon after, it was demonstrated that a single stage isolator would be sufficient to meet those requirements. The conceptual design of the single stage system was inspired by the JILA pre-prototypes [50] and the Stanford experience with the technical demonstrator [51]. The tests performed with the Tech-Demo showed that reasonably stiff suspensions greatly simplify the commissioning process without compromising the isolation performance achievable actively at low frequency. As a consequence, several concepts tested and validated on the Tech-Demo have been used for the design of the HAM-ISI: triangular steel blades for the vertical isolation, flexure rods for the horizontal isolation, non-contact magnetic actuators for the drive, and customized commercial instruments for sensing. Detailed design specifications were written in 2006 [52]. The realization of the structural design and the fabrication and assembly of the first unit was carried out at High

Precision Devices in 2007 [53]. The preliminary test results obtained with this first unit were conclusive. In the following year, each of the two LIGO observatories was equipped with a HAM-ISI unit. They were installed in ultra-high vacuum chambers, and used for the Enhanced LIGO program (an intermediate step between Initial LIGO and Advanced LIGO). Extensive testing was carried out. Results obtained with the two prototypes were presented in Kissel's thesis [54]. During the following years, minor but very useful design adjustments were implemented in the Advanced LIGO version of the HAM-ISI. The revision included design changes to facilitate the production, assembly and commissioning. The inertial sensors were modified with stiffer flexures to eliminate the need for a remote locking system, increasing the reliability of the system [55]. Electrical copper shields were installed on the capacitive position sensor cables to contain the electromagnetic radiation from the modulation drive. A set of six geophones (L4C) were added onto the base of the units used in the interferometer's output mode cleaner which required additional performance. This set of geophones is used to improve the isolation in the band around 10 Hz via feedforward control. Fifteen platforms have been produced for the Advanced LIGO program based on this improved design. The following sections summarize the design and gives an overview of the mechanical features and instrumentation. It presents the control strategy, summarizes the test results and presents the isolation performance obtained with the system.

### 5.2 HAM-ISI architecture and instrumentation

The HAM-ISI is a six-axis passive-active ultra-high-vacuum compatible isolator. The platform can carry more than 500 kg of payload and position it with nanometer resolution.

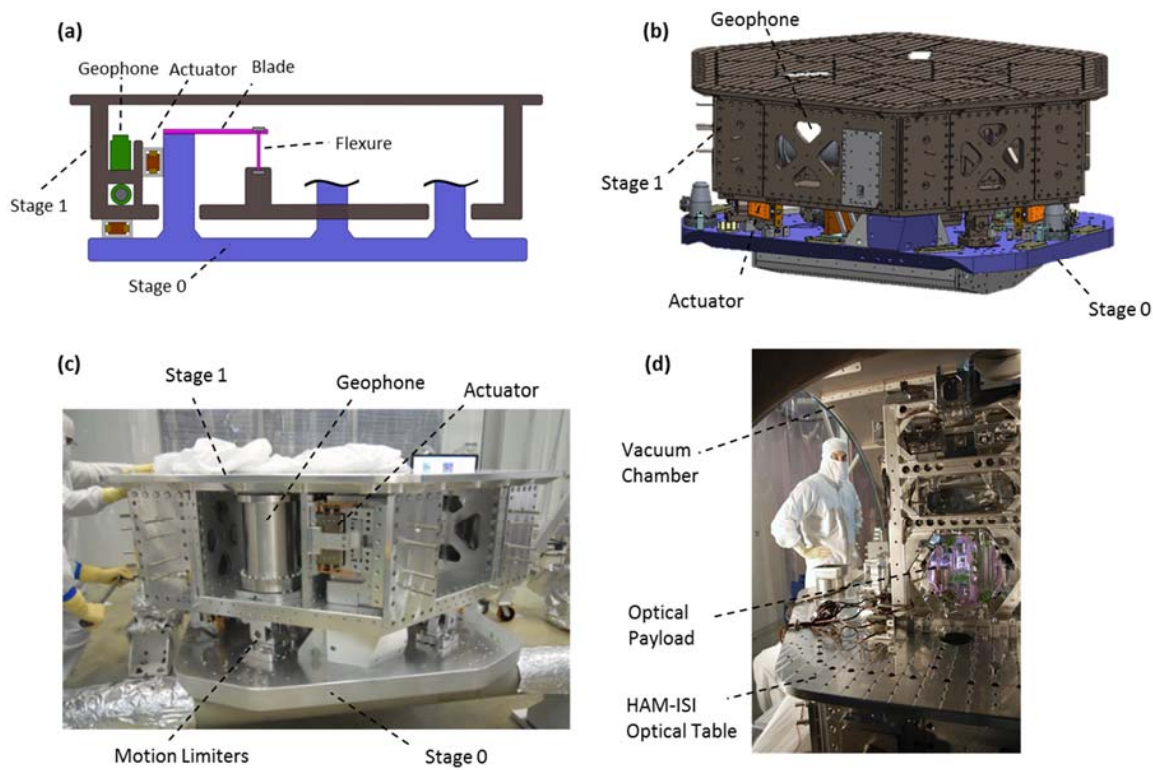

Fig. 13: (a) Conceptual representation of the HAM-ISI, (b) CAD model, (c) picture during assembly, and (d) optical table populated with Advanced LIGO optics.

In Fig. 13 (a), the base of the system is called Stage 0 and the suspended structure is called Stage 1. The top plate of Stage 1 features a 2-m wide hexagonal optical table. Three triangular blades are used to provide the vertical isolation. Three flexure rods are used to provide the horizontal isolation. Blades and flexures are paired to form a spring assembly. Six capacitive position sensors sense the relative motion between the two stages. Six geophones (*Geotech GS13*) are mounted on Stage 1. Six magnetic actuators supplied by *Precision Systems Instruments* are used for the drive between the two stages. Springs, sensors and actuators are divided into three groups positioned symmetrically, 120 degrees apart (only one of them being displayed Fig. 13 (a) for clarity). Six geophones (*Sercel L4C*) are mounted on Stage 0 (HAM4 and HAM5 chambers only). Each group contains one blade-flexure assembly, two actuators and two sensors of each type. Half the instruments are mounted tangentially; the other half are mounted vertically. The sensor noise curves are presented in [56]. Characterization of the voice coil actuators is presented in [57].

The system is designed to operate in an ultra-high vacuum (UHV). Many of the design features are due to this stringent constraint. The bolted structure is made of aluminum 6061 T6, machined with a good finish to help with the cleaning process. All the holes are vented. All the bolting hardware used in the assembly is made of 304 stainless steel (silver plated in all instances for which the receiving threaded hole is also made of steel). Maraging steel is used for the springs and flexures. All the mechanical parts are subjected to a thorough cleaning process, involving chemical cleaning, an ultrasonic bath, and a low-temperature bake process. A combination of residual gas analysis and Fourier transform infrared spectroscopy techniques are used to validate the cleaning of the parts. The capacitive position sensors and magnetic actuators are UHV compatible and are subjected to a similar cleaning process. The geophones are podded in sealed chambers. Fig. 13 (c), shows a pod containing a vertical geophone. The welded pod structures are checked for leaks via a helium test. During the final assembly, the pod is filled with neon. The pod assembly is leak-checked with a residual gas analysis. During operations, the neon is used as tracer of a potential pod leak. Pressure sensors are used in every pod to monitor their internal pressure.

The systems are assembled using standard Class 100 techniques in a dedicated building. After assembly, the system is loaded with a dummy payload to undergo a thorough testing process. To account for any small asymmetries in the vertical spring stiffness, the platform is leveled with balancing masses positioned on the side walls. The testing procedure includes static tests, dynamic tests, cross coupling measurements, sensor noise measurements, open loop and closed loop measurements. Results were repeatable from one platform to another. Once the assembly is validated, the platform is transported to the observatory. It is installed in chamber, mounted on a HEPI, and populated with the optical payload. Fig. 13 (d), shows a HAM-ISI system installed in its chamber and supporting an optic suspended from a triple suspension. After a final round of testing, the vacuum chamber is closed and evacuated. The isolation controllers are then installed and tuned to provide optimal active isolation.

### 5.3 Blades and flexures

Blades and flexures are used to provide the system's compliance in all directions of translation and rotation. The blades provide the vertical flexibility, and the flexure rods provide the horizontal flexibility. These flexible components are designed and positioned such as all six rigid body modes are aligned with the axis of the Cartesian basis.

The triangular blades are mounted horizontally and support the static load Fig. 14 (a). The blades' length, base width and thickness are chosen to satisfy the stiffness and stress targets. Bernoulli beam equations are used in Eq (12) to describe the cross section bending equilibrium. The torque in the cross section varies linearly with $x$, as does the area quadratic moment given in Eq. (13), which gives constant strain and stress as a function of $x$. The stress and deflection are given in Eq. (14) and (15). The vertical stiffness for a load applied at the tip of a blade is given by Eq. (16). The un-deformed profile of the blade is approximated with a constant radius of curvature chosen to obtain a flat blade at full load. Detailed calculations are given in [53].

The deflection, stiffness and stress were checked with non-linear analysis to account for large displacements [58]. The blades geometry and the non-linear FEA results obtained for the spring stiffness, peak stress and maximum deflection are summarized in [53]. The stress profile is shown in Fig. 14 (b). Finite element and experimental modal analysis were done to identify the blades modal shapes. Tuned mass dampers have been designed to damp the blades mode [59].

$$EI \, \frac{\partial^2 w}{\partial x^2} = P \, (l - x) \qquad (12)$$

$$I = \frac{b \, h^3}{12} (l - x) \qquad (13)$$

$$\sigma = \frac{6 \, P \, l}{b_0 \, h^2} \qquad (14)$$

$$w = \frac{6 \, P \, l}{E \, b_0 \, h^3} \, x^2 \qquad (15)$$

$$k_z = \frac{E \, b_0 \, h^3}{6 \, l^3} \qquad (16)$$

Flexure rods are used to provide horizontal flexibility. Fig. 14 (c) shows the translations ($u_0$ and $u_1$), the rotations ($\theta_0$ and $\theta_1$), the forces ($F_0$ and $F_1$), and the moments ($\tau_0$ and $\tau_1$) at the tips of the rod. Beam equations are used to find the location of the forces that will produce pure translation ($\theta_0 = \theta_1 = 0$). Equation (17) gives the static equilibrium of the cross section under bending and normal load $P$, where $E$ is the Young's modulus of the rod, and $I$ is the quadratic moment of area. The general solution is given by Eq. (18), where the constants $a_0$ to $a_4$ are functions of the boundary conditions. Equation (18) can then be expressed as a function of the external forces and substituted into Eq. (17) and its derivatives to produce the stiffness matrix given in Eq. (19). Finally, Eq. (19) can be used to calculate the location ($\lambda$) where the forces $F_0$ and $F_1$ will produce pure translation. This is done by solving Eq. (19) for $\theta_0 = \theta_1 = 0$, which yields to the solution given in Eq. (20). More details regarding these calculations can be found in [60]. The location $\lambda$ is also the called the zero-moment-point location since the sum of the moments of all external forces ($P$, $-P$, $F_1$ and $R_0$) is null at this point. The zero-moment-point distance is used to define the location of the rod relative to the blade and the actuator. The actuator is positioned at a distance $\lambda$ from the bottom tip so that the actuator force $F_1$ will produce a pure translation. The blade is positioned at a distance $\lambda$ from the top tip, so that the reaction force from the blade will not rotate the rod. Equation (21) gives an approximation of the stress ($\sigma$) as a function of the beam deflection. The first term gives the axial stress and the second term gives the bending stress where $\beta$ is a stress concentration factor. Equation (22) gives the bending moment as a function of the lateral deflection (assuming $u_0 = 0$). The maximum lateral deflection is limited to 0.4 mm by motion limiters. These equations are used to design the rod as a function of the desired lateral stiffness, lateral deflection and stress. The results have been checked via finite element analysis and measurement [53].

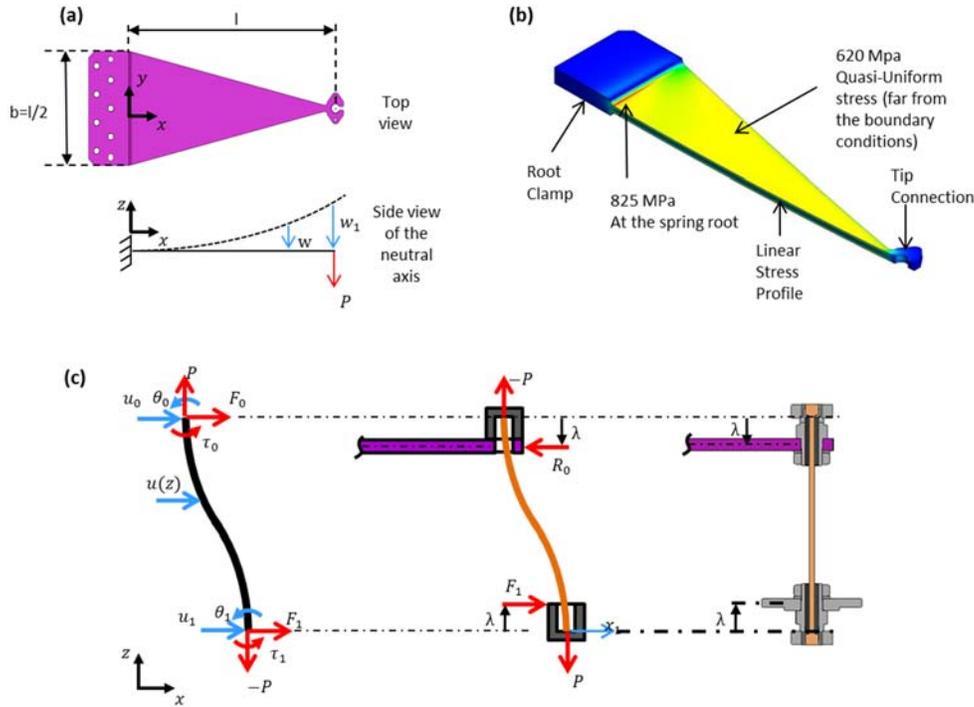

**Fig. 14: Triangular blades**

$$EI \frac{\partial^2 u}{\partial z^2} = -\tau_1 + F_1 z + P(u - u_1) \tag{17}$$

$$u(z) = a_0 + a_1 z + a_2 \cosh(c z) + a_3 \sinh(c z) \tag{18}$$

$$\begin{Bmatrix} F_0 \\ \tau_0 \\ F_1 \\ \tau_1 \end{Bmatrix} = \begin{bmatrix} k_{tt} & k_{tr} & -k_{tt} & k_{tr} \\ k_{tr} & k_{rr} & -k_{tr} & k'_{rr} \\ -k_{tt} & -k_{tr} & k_{tt} & -k_{tr} \\ k_{tr} & k'_{rr} & -k_{tr} & k_{rr} \end{bmatrix} \begin{Bmatrix} u_0 \\ \theta_0 \\ u_1 \\ \theta_1 \end{Bmatrix} \tag{19}$$

$$\lambda = \frac{k_{tr}}{k_{tt}} = \sqrt{\frac{EI}{P}} \tanh\left(\frac{L}{2}\sqrt{\frac{P}{EI}}\right) \tag{20}$$

$$\sigma = \frac{P}{S} + \beta \frac{M d}{2 I} \tag{21}$$

$$M = \frac{\lambda P}{L - 2\lambda} u_1 \tag{22}$$

## 5.4 HAM-ISI control and experimental results

Fig. 15 (a) shows the HAM-ISI transfer functions for the horizontal (X), vertical (Z), pitch (RY) and yaw (RZ) DOF. For each of the curves, the actuators are combined to drive the platform along the Cartesian basis and the inertial sensors are used to sense the motion in the same direction. The transfer functions show that the platform behave as single DOF rigid body system in all directions up to 100 Hz. Repeatable results were obtained for all the platforms. Mass, stiffness and natural frequencies are summarized in Table 1, and more details can be found in [53].

**Table 1: HAM-ISI natural frequencies, mass, inertia and stiffness properties**

| Direction | Mass/Inertia | Stiffness | Frequency |
| --- | --- | --- | --- |
| X, Y | 1973 kg | $1.31 \times 10^5$ N/m | 1.3 Hz |
| Z | 1973 kg | $2.43 \times 10^5$ N/m | 1.8 Hz |
| RX, RY | 530 kg.m$^2$ | $2.42 \times 10^4$ N.m/rad | 1.1 Hz |
| RZ | 825 kg.m$^2$ | $2.65 \times 10^4$ N.m/rad | 0.9 Hz |

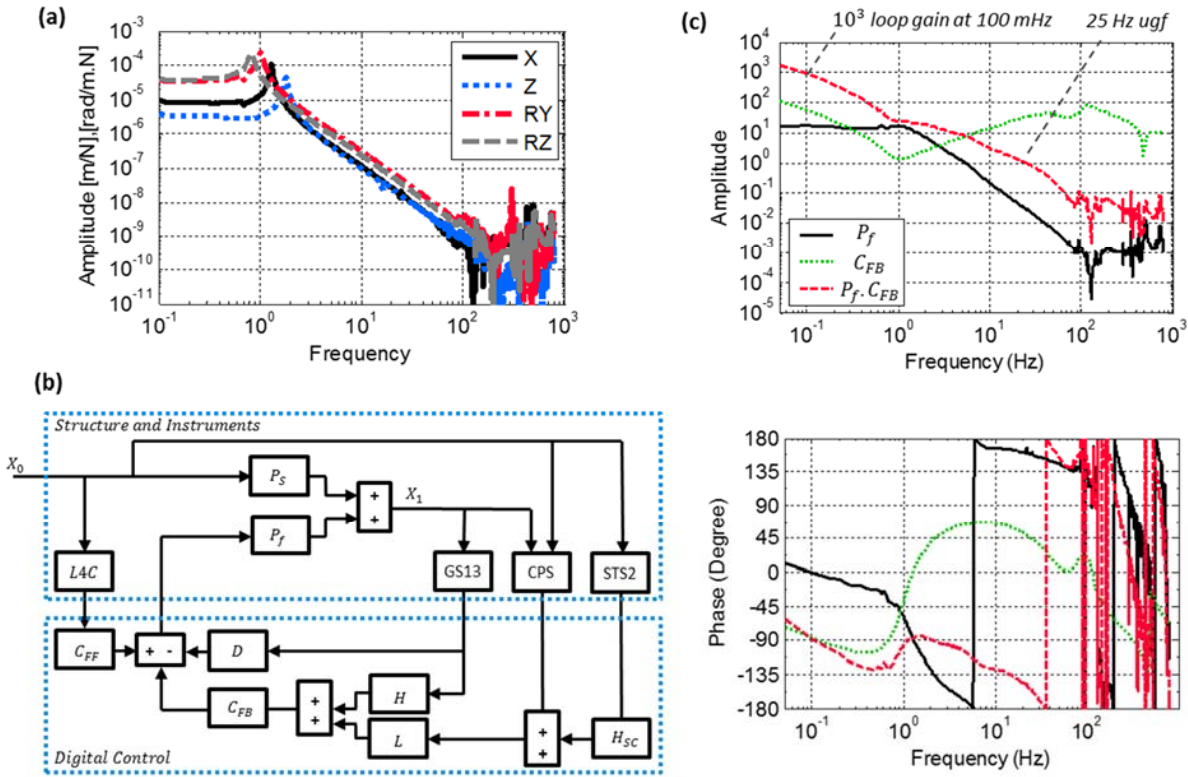

**Fig. 15: (a) HAM-ISI transfer functions, (b) control diagram and (c) example of controller**

The control diagram of the HAM-ISI system is shown in Fig. 15 (b). In addition to the general features described in section 3, a damping filter $D$ is used to reduce the rigid body modes resonances shown in Fig. 15 (a). This controller reduces the risk of saturation during the phases of testing and commissioning (before the more sophisticated isolation loops are designed and installed). The inertial and relative sensor signals are blended using the complementary filters $H$ and $L$ as explained in section 3. These two filters are specifically tuned for the HAM-ISI instruments characteristics. The isolation filter is fed with the resulting blended signal. Fig. 15 (c) shows an example of a HAM-ISI control loop. The solid curve shows the plant transfer function, the dotted line shows the isolation filter and the dashed line shows the open-loop transfer function. In this example, the unity gain frequency is 25 Hz and the phase margin is 30 degrees. The loop gain is approximately 1000 at 100 mHz. Additionally, low-noise ground instruments (*STS2*) are used for sensor correction in the 0.1 Hz to 1 Hz band, through the high pass filter $H_{sc}$. In the 10 Hz region, *L4Cs* mounted on Stage 0 of HAM4 and HAM5 are used for feedforward control ($C_{FF}$) to suppress the residual seismic noise.

Fig. 16 and Fig. 17 show typical isolation results obtained with the HAM-ISI system. The ground motion is measured with a long-period seismometer installed near the vacuum chambers. The platforms motion is measured with in-loop inertial sensors (GS13). Fig. 16 (a) show that the HAM-ISI motion in the X and Y direction which directly couple to the length of the optical cavities [61]. The curves show that the system meet the requirements at all frequencies above 0.5 Hz. The ability to improve the performance below 0.5 Hz is limited by the tilt horizontal coupling. In the frequency band of the suspension modes (above 0.5 Hz), the HAM-ISI performance exceeds the requirements, and meet the sensor noise at 1 Hz.

Fig. 16 (b) shows that the motion in the X direction of all the HAM-ISI units of a detector. The curves show that all the platforms have very similar performance at low frequencies. Above 10 Hz, the spectral densities show slightly different features related to the mechanical resonances of the support structures. These features are heavily filtered by the passive suspensions of the auxiliary optics.

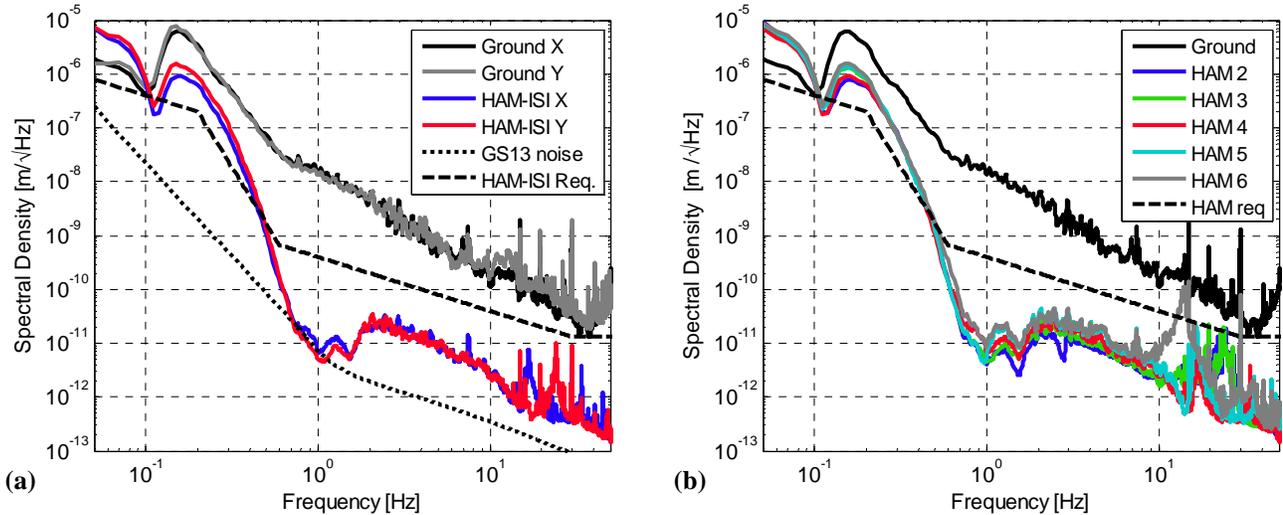

**Fig. 16. (a) HAM-ISI isolation performance for the horizontal DOF, and (b) HAM-ISI isolation in the X direction for all of the 5 units of a detector.**

Fig. 17 (a) estimates the longitudinal motion of the suspension point of an auxiliary optic (the longitudinal direction is normal to the optics face). This optic is oriented with a large angle with respect to the HAM-ISI axis, therefore it is sensitive to both the X and Y motion of the HAM-ISI. The optic off center of the optical table, therefore the longitudinal motion of the suspension point is also sensitive to the RZ motion of the HAM-ISI. Finally, the suspension point of this optic is located 0.75 meters above the optical table, therefore the longitudinal motion of the suspension point is also sensitive to the RX and RY motion of the HAM-ISI. This example illustrates why the seismic isolation systems must provide isolation in all directions of translation and rotation. Fig. 17 (a) shows how each degree of freedom of the HAM-ISI contributes to the longitudinal motion of the optic. Below 0.5 Hz, the longitudinal motion of the optic is dominated by X and Y motion of the HAM-ISI. Above 0.5 Hz, the longitudinal motion of the optic is dominated by the RX and RY contribution of the HAM-ISI.

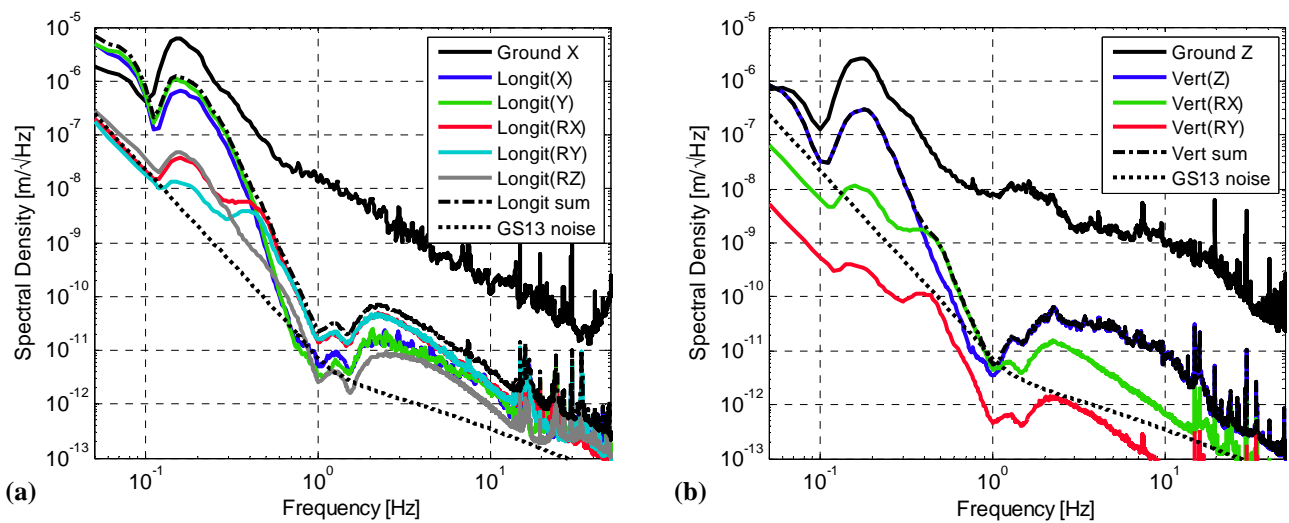

**Fig. 17. Plots showing the contribution of the HAM-ISI DOF to: (a) the longitudinal motion of the suspension point and (b) the vertical motion of the suspension point.**

Fig. 17 (b) estimates the contribution of the HAM-ISI DOF to the vertical motion of the suspension point which must be well isolated as it strongly couples into pitch motion of the optic. The vertical motion of the suspension point is limited by the HAM-ISI rotation between 0.5 Hz and 1 Hz. Further improving the performance in this frequency band is possible at the cost of deteriorating the performance at lower frequencies through sensor noise injection and tilt-coupling.

Beyond isolation performance, the past years of commissioning of the two Advanced LIGO interferometers showed that the HAM-ISI robustly supports the operations of the detectors. The next section gives an overview of the instrumentation and performance of the BSC-ISI system used for the interferometers core optics.

## 6 The two-stage internal isolator (BSC-ISI)

### 6.1 BSC-ISI Overview

The seismic motion requirements for the Advanced LIGO core optics were defined in 1999. The optics require an isolation factor of 10 at 0.1 Hz, and up to several thousands in the control bandwidth [47]. To provide such isolation, an active isolation concept inspired by the JILA pre-prototype was proposed [6]. A Rapid Prototype, shown in Fig. 18 (a), was constructed. The mechanical design was done at High Precision Devices. The system was built and tested at MIT [15]. The results of the testing demonstrated both performance and robustness which motived the stiff suspension concept to become the baseline design for Advanced LIGO [13].

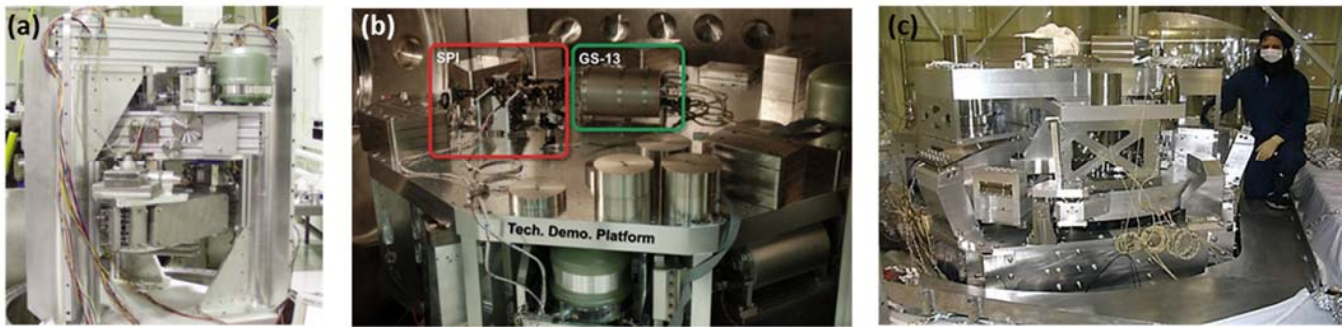

**Fig. 18. (a) the two-stage Rapid Prototype, (b) the two-stage Technical Demonstrator, and (c) the BSC-ISI prototype.**

A technical demonstrator (Tech-Demo), Fig. 18 (b), was built during the following years [51]. The mechanical design was realized by High Precision Devices. The system was tested at Stanford. The results obtained with the Tech-Demo validated the choice of the two-stage system. The decision was made that a two-stage system (the BSC-ISI) would be designed and used for Advanced LIGO. The detailed specifications for the BSC-ISI design were defined in 2003 [62]. The realization of the BSC-ISI prototype design was carried out by Alliance Space systems Inc. [63, 64]. Between 2006 and 2008, a prototype, Fig. 18 (c), was assembled and tested at the LIGO MIT test facility [65]. The tests carried out proved the concept. The design was then re-engineered from 2009 to 2011 to incorporate the lessons learned during the prototyping phase and to make it suitable for the timely production of the 15 units needed for Advanced LIGO [66]. The final design was successfully tested at MIT in 2011. In the past two years, 15 platforms have been assembled and tested at the LIGO sites. A detailed presentation of the final design and production process aspects is given in ref. [67], and a detailed presentation of the testing process and experimental results is given in ref. [68].The following section gives an overview of the two-stage system design. It presents the control strategy along with the isolation performance achieved with the platform.

### 6.2 BSC-ISI architecture and instrumentation

A schematic representation of the BSC-ISI system is shown in Fig. 19 (a). The base of the system (Stage 0) supports the first suspended stage (Stage 1) with three sets of blades and flexures. The output stage (Stage 2) is supported from Stage 1 using a similar set of blades and flexures. Stage 1 is instrumented with 6 relative capacitive position sensors, 3 three-axis Nanometric Trillium T240 seismometers, and 6 Sarcelles L4C geophones. Stage 2 is instrumented with 6 relative capacitive position sensors and 6 Geotech GS13 geophones.

The inertial sensors are podded for ultra-high-vacuum compatibility as described in the previous section for the HAM-ISI system. Mu-metal shields were added in the geophone pods to reduce the coupling between the magnetic actuator and the sensor. Triaxal cables are used for the capacitive position sensors to contain the electromagnetic radiation.

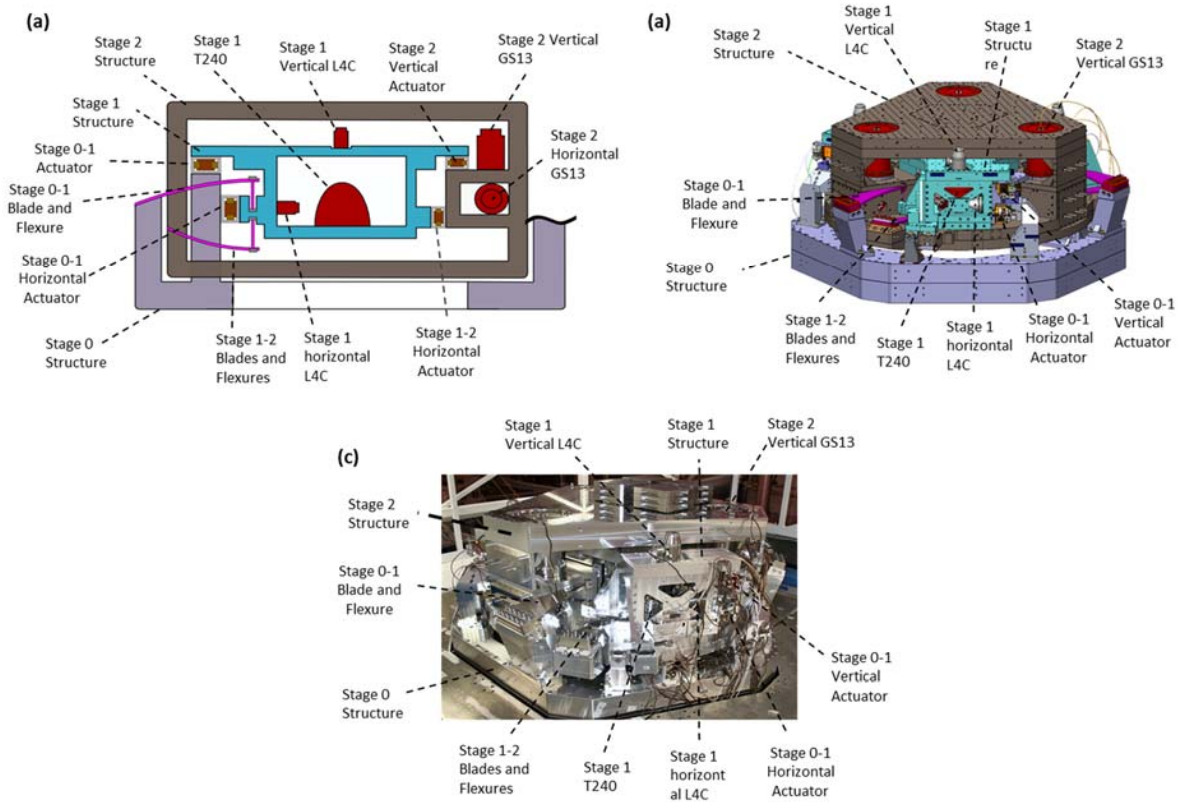

**Fig. 19: (a) schematic representation of the BSC-ISI system, (b) CAD model, and (c) picture of one of the fifteen assemblies.**

Three blade-flexure assemblies are used on each stage and positioned symmetrically at 120 degrees around the vertical axis. The concepts presented in section 4 for the HAM-ISI blades and flexures are also used for the BSC-ISI. The main difference is that the BSC-ISI blades are flat in the un-deformed state. They are curved with a constant radius of curvature in the loaded configuration. This option considerably reduces the loss (and cost) of bulk material during the machining process. For compactness, the stage 0-1 blade center of curvature is located under the blade (with respect to the vertical axis), while the stage 1-2 blade center of curvature is located above the blade. Flexure rods similar to those described previously for the HAM-ISI are used for both stages. Details of the BSC-ISI springs' analysis can be found in [63, 69, 70]. Table 2 summarizes the stiffness and modal properties of the BSC-ISI system.

**Table 2: Inertia, stiffness and natural frequencies of the BSC-ISI system**

| Direction | Stage 0-1 Stiffness | Stage 1-2 Stiffness | In Phase Mode | Out of Phase Mode |
| --- | --- | --- | --- | --- |
| X (or Y) | $2.73 \times 10^5$ N/m | $5.76 \times 10^6$ N/m | 1.25 Hz | 5.22 Hz |
| Z | $6.83 \times 10^5$ N/m | $8.26 \times 10^6$ N/m | 1.82 Hz | 6.68 Hz |
| RX (or RY) | $1.84 \times 10^4$ N.m/rad | $2.18 \times 10^3$ N.m/rad | 1.55 Hz | 6.78 Hz |
| RZ | $1.44 \times 10^3$ N.m/rad | $2.95 \times 10^3$ N.m/rad | 1.42 Hz | 5.68 Hz |

### 6.3 BSC-ISI control and experimental results

Fig. 20 (a) shows the control topology of the BSC-ISI system. All twelve DOF are controlled independently in the Cartesian basis. The long-period three-axis seismometers (*T240*) of the first stage provide the low noise performance necessary at low frequency. Positioning these instruments on stage 1 allows them to be used both as feedback sensors on stage 1 and feedforward sensors for stage 2 through a sensor correction scheme. The *T240*, *L4C* and *CPS* signals of stage 1 are combined with the *H*, *M*, and *L* blend filters. The filter *L* low-passes the *CPS* signal to provide positioning capability at low frequency and filters it at higher frequency to permit seismic isolation. The filters *M* and *H* combine the *T240* and *L4C* signals to provide a very low noise and broadband inertial sensing combination. At low frequency, they filter the inertial signals that are sensor noise dominated (or tilt dominated for the horizontal inertial sensors). Feedforward and sensor correction paths are included in the control scheme as discussed for HEPI and the HAM-ISI. The control diagram of Stage 2 is similar to the HAM-ISI.

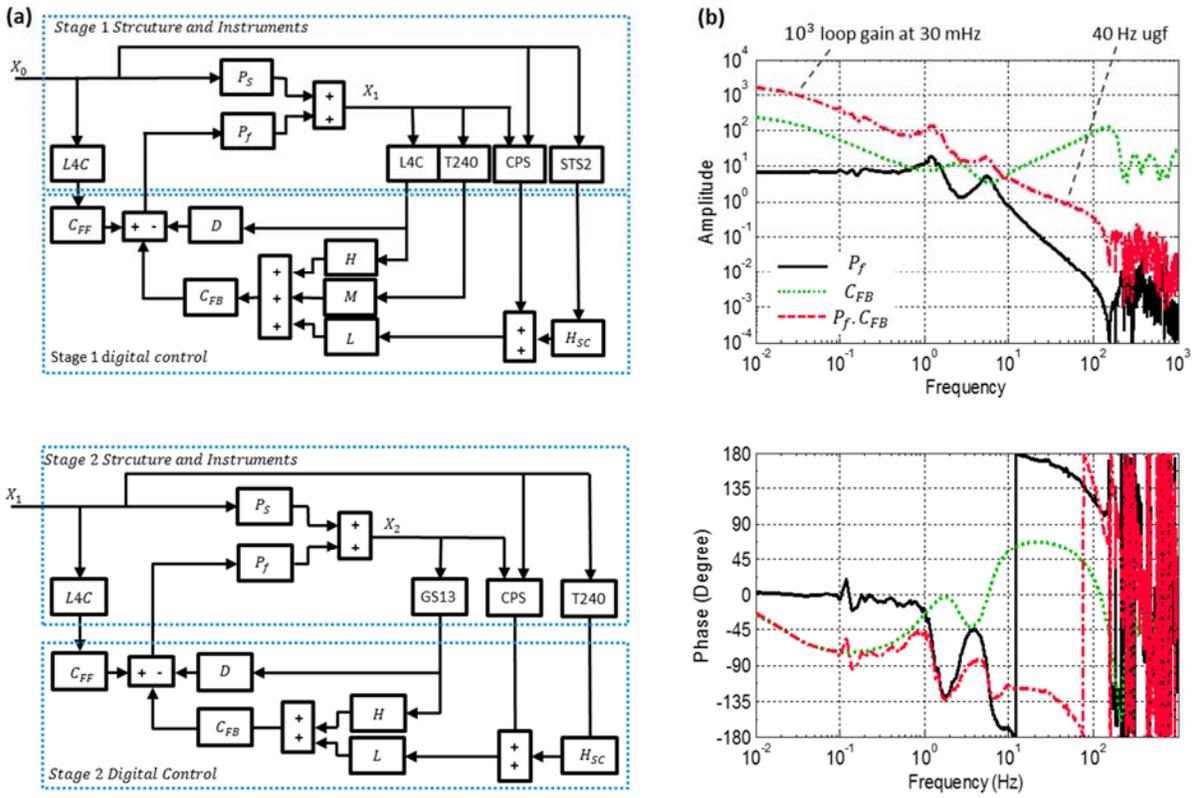

**Fig. 20. Example of BSC-ISI transfer functions.**

An example of controller is shown in Fig. 20 (b) for the X degree of freedom of Stage 1. The solid line shows the plant transfer function. The two rigid body modes are damped by the active damping filters *D*. Above 150 Hz, the peaks of the structural resonances damped with passive devices [66, 71]. The compensator shown by the dotted line is designed to provide high bandwidth (40 Hz unity gain frequency). The open loop shown by the dashed line has 45 degrees of phase margin and provides high loop gain in the control bandwidth.

Fig. 21 and Fig. 22 show typical isolation results obtained with the BSC-ISI system. The ground motion is measured with a long-period seismometer installed near the vacuum chambers. The platform's motion is measured with the in-loop inertial sensors of Stage 2 (GS13).

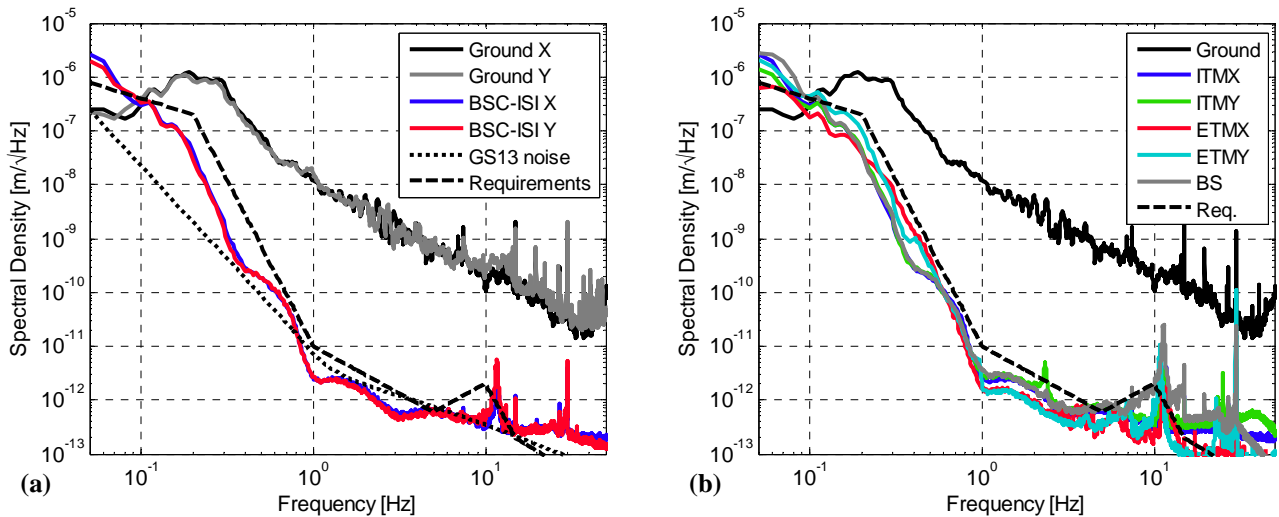

**Fig. 21. Typical BSC-ISI isolation performance: (a) horizontal DOF, and (b) Comparison of the performance of the 5 units.**

Fig. 21 (a) show that the BSC-ISI motion in the X and Y direction which directly couples to the 4 km arms cavities and to the Michelson length. The curves show that the system meets the requirements at all frequencies from 0.1 Hz to 10 Hz. The performance is limited by tilt horizontal coupling at low frequencies. Between 1 Hz and 5 Hz, the in loop measurement appears below the theoretical noise of the instrument, indicating that the actual motion is completely sensor noise limited. Above 10 Hz the performance is loop gain limited.

Fig. 21 (b) shows that the motion in the X direction of all the BSC-ISI units of a detector. The curves show that all the platform have very similar performance at all frequencies up to 10 Hz. The higher frequencies features are due to structural and instrument noise differences between the units and their support structure. These features are strongly filtered by the passive suspensions supporting the core optics.

Fig. 22 (a) estimates the longitudinal motion of the suspension point of a quadruple pendulum, where the longitudinal direction is normal to the optics face. Since the mirror is aligned with X axis of the arm cavity, there is no theoretical contribution of the Y and RX degrees of freedom of the BSC-ISI. Below 0.25 Hz, the longitudinal motion of the optic is dominated by the X motion of the BSC-ISI. Above that frequency, the RY DOF significantly contribute.

Fig. 22 (b) estimates the contribution of the BSC-ISI DOF to the vertical motion of the suspension point. The performance around 0.5 Hz is particularly important as the vertical motion of the quadruple pendulum couples into pitch modes of the test masses which are particularly difficult to control. At those frequencies, the motion is limited by the rotational DOF. Further performance improvement is limited by the sensor noise of the vertical inertial sensors, which amplifies the low frequency tilt motion. This results in horizontal motion amplification through tilt-horizontal coupling in the horizontal inertial sensors. Thanks to the very high level of isolation provided the BSC-ISI at low frequencies, the detector operate robustly with a high duty cycle.

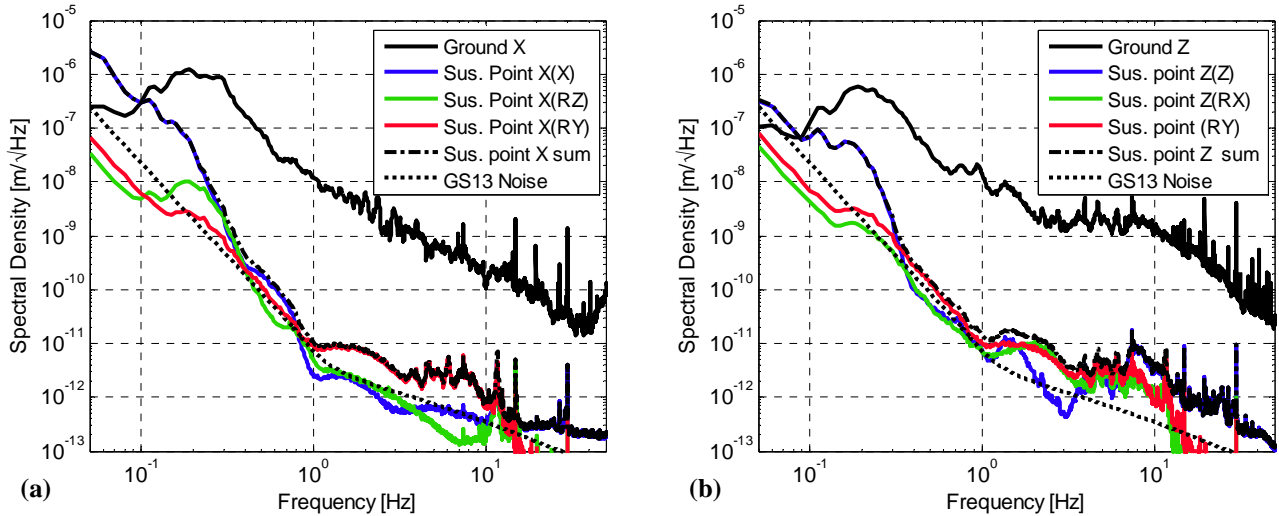

**Fig. 22. Plots showing the contribution of the BSC-ISI DOF to: (a) the longitudinal motion of the suspension point and (b) the vertical motion of the suspension point.**

## 7 Conclusion

This paper has reviewed a decade of research on seismic isolation for Advanced LIGO gravitational wave detectors. An overview of the system's strategy developed to meet Advanced LIGO requirements was given. The stages of isolation are composed of two major categories: active-passive platforms which provide low frequency isolation using very low noise inertial sensors, and mostly passive suspensions which provide high frequency isolation. This strategy greatly facilitated the integration of complex, interdependent subsystems as Advanced LIGO was installed and commissioned.

We presented the detailed history that led to the final design for the three types of active-passive seismic isolation platforms used in Advanced LIGO. For each system, a detailed presentation of the mechanical design, instrumentation, control strategy and isolation performance was given. At each of the two LIGO observatories, 11 pre-isolators, 5 single stage internal isolators, and 5 two-stage isolators have been constructed, installed, populated with suspensions and commissioned in less than four years. In addition to the 42 platforms installed at the Livingston and Hanford sites, another set of 21 isolators was produced for the third Advanced LIGO interferometer which we anticipate to be installed in India.

Test isolation results summarized in this review show that the seismic isolation platforms meet the very stringent motion requirements that were defined for Advanced LIGO more than a decade ago. The past two years of interferometric commissioning have also proven the seismic isolation design to be very robust and effective. The isolators currently support long, stable stretches of full interferometer resonance as the Advanced LIGO detectors prepare for observation runs.

## Acknowledgments

The authors gratefully thank the National Science Foundation for their support. LIGO was constructed by the California Institute of Technology and the Massachusetts Institute of Technology with funding from the National Science Foundation and operates under cooperative agreement PHY-0757058. Advanced LIGO was built under award PHY-0823459. We thank the JILA group for pioneering the work on active isolation systems using low frequency inertial sensors, and for demonstrating the feasibility of such multi-stage systems. We thank our colleagues from the suspension groups in GEO, VIRGO and LIGO for introducing us to the benefits of using triangular Maraging blades to provide vertical isolation. We thank High Precision Devices for the mechanical design's realization of the Rapid Prototype, the technical demonstrator, and the single stage isolator. We thank Alliance Space systems Incorporation for the mechanical design's realization of the two-stage prototype. We thank Nanometrics, Streckeisen, Geotech, Sarcelles and Microsense for supplying us with great instruments, and for their technical support. The drawings in Fig. 9 (a), Fig. 13 (a) and Fig. 19 (a) are to the credit of Caleb Johns. Figure 9(c), 13(b)(c) and 18(c) were previously used in our conference papers published in the proceedings of the ASPE, [67] and [74], and are re-used with permission from the ASPE. Figure 13 (d) is to the credit and is a courtesy of Kate Gushwa, Matthew Heinze and the LIGO Suspension group. Figure 14 re-use content from LIGO and HPD internal documents [55] and [60]. Figure 19(c) was previously published in our conference paper on the dynamics enhancements of the two-stage system [68], and is re-used with permission from the ASME. Figure 20 combines figures previously used in our article on the BSC-ISI two-stage system, published in Precision Engineering [70]. The figures are re-used and combined with permission from Elsevier. Finally yet importantly, this work would not have been possible without the outstanding support of the LIGO laboratory management, computer and data systems, procurement, facility modification and preparation, assembly and installation teams. This document has been assigned LIGO Laboratory document number LIGO-P1200040. Public internal LIGO documents are found at: https://dcc.ligo.org/cgi-bin/DocDB/DocumentDatabase/.

## Authors contribution